\documentclass[12pt]{article}
\usepackage{amsmath,amssymb,amsfonts,amscd,bbm,graphicx,cite}
\begin{document}


\thispagestyle{empty}
\renewcommand{\thefootnote}{\fnsymbol{footnote}}
\setcounter{footnote}{1}

\vspace*{-1.cm}
\begin{flushright}
OSU-HEP-05-06
\end{flushright}
\vspace*{1.8cm}

\centerline{\Large\bf Quark and Lepton Masses}
\vspace*{3mm}
\centerline{\Large\bf from Deconstruction}

\vspace*{18mm}

\centerline{\large\bf T. Enkhbat\footnote{E-mail:
\texttt{enkhbat@okstate.edu}} and G. Seidl\footnote{E-mail:
\texttt{gerhart.seidl@okstate.edu}}}

\vspace*{5mm}
\begin{center}
{\em Department of Physics, Oklahoma State University}\\
{\em Stillwater, OK 74078, USA}
\end{center}

\vspace*{20mm}

\centerline{\bf Abstract} \vspace*{2mm} We propose a
supersymmetric $SU(5)'\times SU(5)''$ model, where the quarks and
leptons live in a $U(1)$ product group theory space that is compactified
on the real projective plane $RP^2$. The fermion generations are placed
on different points in the deconstructed manifold by assigning
them $SO(10)$ compatible $U(1)$ charges. The observed masses and
mixing angles of quarks and leptons emerge from non--renormalizable
operators involving the chiral link fields. The link fields
introduce a large atmospheric neutrino mixing angle
$\theta_{23}\sim 1$ via a dynamical realization of the seesaw
mechanism, which sets the deconstruction scale to a value of the
order the usual $B-L$ breaking scale $\sim 10^{14}\:{\rm GeV}$.
Supersymmetry breaking can be achieved through topological effects
due to a nontrivial first homology group $Z_2$. The mixed
anomalies of the link fields are canceled by Wess--Zumino terms,
which are local polynomials in the gauge and link fields only. We
also comment on the construction of Chern--Simons couplings from
these fields.


\renewcommand{\thefootnote}{\arabic{footnote}}
\setcounter{footnote}{0}

\newpage
\section{Introduction}
There are many reasons why the Standard Model (SM) should be
extended. The chief examples to which the SM does not provide any
answer are the gauge hierarchy problem, charge quantization, and
the origin of fermion masses and mixings. Supersymmetry (SUSY) and
four-dimensional (4D) Grand Unified Theories (GUT's) give partial
solutions to the first two of the above problems but not to the
latter. To understand the observed pattern of fermion masses and
mixing angles, it seems therefore necessary that a new ingredient
must be added, which allows to distinguish between the generations
in a controlled way. For this purpose, one usually advocates a
flavor symmetry. Generally, models based on continuous non-Abelian
flavor symmetries are highly dependent on the details of the
flavor symmetry breaking, without referring to deeper underlying
dynamics. The models using an Abelian flavor symmetry, on the
other hand, have as a common feature that the three generations
carry different charges. At least from a bottom--up point of view,
however, generation--dependent charges seem to be somewhat
contrary to the spirit of GUT's, wherein the ad hoc assignment of
hypercharges to the quarks and leptons is explained.

In recent years, higher--dimensional theories opened up new
possibilities along this direction \cite{Arkani-Hamed:1998rs,
Randall:1999ee}. For example, instead of assuming that the quarks
and leptons carry generation--dependent charges under a flavor
symmetry, the generations might be distinguished by their position
in an extra dimension. A hierarchy of Yukawa couplings could then
arise from the overlap of the spatial wave--functions of the matter
fields in the extra dimension \cite{Arkani-Hamed:1999dc}. It would
now be interesting to simulate or reproduce this
higher--dimensional mechanism in a conventional 4D field theory,
which is manifestly gauge--invariant and renormalizable. This can
be achieved by employing the idea of dynamically generated extra
dimensions, called deconstruction
\cite{Arkani-Hamed:2001ca,Hill:2000mu}. In
deconstruction,\footnote{For an early approach in terms of
infinite arrays of gauge theories, see
Ref.~\cite{Halpern:1975yj}.} one considers the extra dimensions as
an infrared effect of an ultraviolet complete theory described by
a product of 4D gauge groups $\Pi_i \bigotimes G_i$. The
deconstructed dimensions are represented in a ``theory space''
\cite{Arkani-Hamed:2001ed}, where the gauge groups $G_i$
correspond to ``sites'' that are connected by ``links'', like in a
transverse lattice gauge theory \cite{Bardeen:1976tm}. Such a view
of extra dimensions has rich theoretical and phenomenological
implications covering studies in different directions and energy
scales. These studies include, for example, electroweak symmetry
breaking \cite{Arkani-Hamed:2001nc}, GUT-type of models
\cite{Witten:2001bf,Cremades:2002te}, supersymmetry breaking
\cite{Arkani-Hamed:2001ed,Luty:2001jh,Csaki:2001em,Dudas:2003iq},
and fermion masses and mixings
\cite{Kaplan:2001ga,Skiba:2002nx,Balaji:2003st}.\footnote{Deconstruction
has, for example, also been applied to neutrino oscillations
\cite{Hallgren:2004mw}, the Casimir effect \cite{Bauer:2003mh},
instantons \cite{Csaki:2001zx}, gravity
\cite{Arkani-Hamed:2002sp}, and calculable models of the ``landscape'' of
string vacua \cite{Dienes:2004pi}.} Yet, a realistic deconstructed
model, which gives all the observed fermion masses and mixing angles in
the framework of a GUT, has not been proposed so far. This is the
aim of the present paper.

The 4D product GUT's which exhibit a higher-dimensional
correspondence via deconstruction, have the advantage that
dangerous proton decay operators can be easily suppressed by
discrete symmetries. The doublet--triplet splitting problem, for
example, can be solved in a model proposed by Witten
\cite{Witten:2001bf}, which is based on a 4D SUSY $SU(5)$ product
GUT that is obtained from deconstruction. In the present paper, we extend
this model by a $U(1)^N$ theory space. The different generations
of quarks and leptons populate this space and are located at
different sites in such a way, that the fermion  masses and mixings
emerge naturally. A simple linear structure of the product group
space, corresponding to a single extra dimension, seems to be too
restrictive to account for the entire fermion mass and mixing
pattern of the SM. Therefore, we start instead with a
deconstructed two-dimensional disk, which can be part of an even
larger structure, the so called ``spider web theory space''
introduced in Ref. \cite{Arkani-Hamed:2001ed}. It was shown in
Ref.~\cite{Arkani-Hamed:2001ed}, that when the spider web theory
space is converted into the real projective plane $RP^2$,
supersymmetry breaking can be viewed as arising from a topological
obstruction due to a nontrivial first homology group
$H^1(RP^2)=Z_2$. In spider web theory space, one can therefore
simultaneously account for SUSY breaking and the generation of
fermion masses and mixings.

To ensure the consistency of our model, we have to address the
anomalies associated with the enlarged gauge symmetry in four
dimensions. Anomaly--cancelation in theory space has been previously
discussed in Refs.~\cite{Dudas:2003iq,Skiba:2002nx,Falkowski:2002vc,Giedt:2003xr}.
The cancelation
of the anomalies is generally carried out by introducing
appropriate Wess--Zumino terms \cite{Wess:1971yu}, which represent
non-decoupling effects of heavy fermions in the low--energy theory.
We apply this approach to our spider web theory space. In
addition, we examine the continuum limit of Chern--Simons terms,
which, however, do not contribute to the anomalies.

The paper is organized as follows. In Sec.~\ref{sec:U(1)},
we present our model. In Sec.~\ref{sec:23splitting}, we review the
solution to the doublet--triplet splitting problem in an $SU(5)'\times
SU(5)''$ product GUT. Next, in Sec.~\ref{sec:U1theoryspace}, we
introduce our model for the quark and lepton masses based on a
$U(1)$ spider web theory space. We also comment in this section on
supersymmetry breaking via the nontrivial topology of $RP^2$. The generation
of the fermion masses and mixings is described in
Sec.~\ref{sec:quarksandleptons}. The predictions for the
up--quarks, down--quarks/charged leptons, and neutrinos are presented
in Secs.~\ref{sec:upsector}, \ref{sec:downsector}, and
\ref{sec:neutrinos}. The anomaly cancelation in our model is
discussed in Sec.~\ref{sec:anomalies}. Finally, we give our conclusions in
Sec.~\ref{sec:conclusions}.

\section{Deconstructed $U(1)$}\label{sec:U(1)}
It has been proposed by Witten, that the doublet--triplet splitting
problem can be solved in an $4D$ SUSY $SU(5)'\times SU(5)''$
product GUT model, which arises from deconstruction
\cite{Witten:2001bf} (a similar approach has been given earlier by
Barbieri {\it et al.} \cite{Barbieri:1994jq}). In this section, we
will build upon this setup and extend it to a model, which
reproduces the observed fermion masses and mixings. We will first
begin in Sec.~\ref{sec:23splitting} with a brief review of the
known solution to the doublet--triplet splitting problem, which we
then take in Sec.~\ref{sec:U1theoryspace} as a starting point for
introducing our model of quark and lepton masses.

\subsection{Doublet--triplet splitting in $SU(5)'\times SU(5)''$}\label{sec:23splitting}
Following the doublet--triplet splitting mechanism proposed by
Witten and Barbieri {\it et al.}, one assumes a 4D gauge group
$G=SU(5)'\times SU(5)''$, in which the SM gauge group
$G_{SM}=SU(3)_C\times SU(2)_L\times U(1)_Y$ is embedded as a
diagonal subgroup. The model possesses a discrete global symmetry
$F=Z_N$ which commutes with $G$. At the GUT scale, the symmetry
group $G\times F$ is spontaneously broken down to $G_{SM}\times
F'$, where $F'$ is a linear combination of $F$ and the $Z_N$
subgroup of the hypercharge subgroup $U(1)_Y''$ of $SU(5)''$. The
MSSM Higgs doublets are contained in the $SU(5)'\times SU(5)''$
representations
\begin{equation}\label{eq:Higgses}
 ({\bf 5},{\bf 1})^H+({\bf 1},\overline{\bf 5})^H,
\end{equation}
{\it i.e.}, the Higgs superfield that gives masses to the up
quarks transforms under the fundamental representation of $SU(5)'$
and is a singlet under $SU(5)''$. The Higgs which generates the
down quark and charged lepton masses, on the other hand, is in the
antifundamental representation of $SU(5)''$  and is an $SU(5)'$
singlet. Under $G\supset G_{SM}$, the Higgs fields in
Eq.~(\ref{eq:Higgses}) decompose as ${\bf 5}^H=(Q,H)$ and
$\overline{\bf 5}^H= (\tilde{Q},\tilde{H})$, in which $H$ and
$\tilde{H}$ are the MSSM Higgs doublets, whereas $Q$ and
$\tilde{Q}$ are their corresponding color triplet partners. The
crucial point which now allows to solve the doublet--triplet
splitting problem is here that the unbroken discrete symmetry $F'$
commutes with $SU(5)'$ but not with $SU(5)''$. As a result, $F'$
acts on the whole multiplet ${\bf 5}^H$ but distinguishes in
$\overline{\bf 5}^H$ between the triplet and doublet  components
$\tilde{Q}$ and $\tilde{H}$. One can therefore have an
$F'$-invariant coupling $\tilde{Q}Q$ in the superpotential while a
$\mu$-term-type coupling $\sim H\tilde{H}$ is (at the GUT scale)
forbidden by $F'$, which solves the doublet--triplet splitting
problem.

When including quarks and leptons in this model, it is necessary that $F'$ can
forbid all dangerous baryon number violating operators, which would otherwise
mediate proton decay. This becomes indeed possible
\cite{Witten:2001bf}, when we assume that under $SU(5)'\times SU(5)''$ the
matter superfields transform as
\begin{equation}\label{eq:fermions}
({\bf 10},{\bf 1})_i+(\overline{\bf 5},{\bf 1})_i+({\bf 1},{\bf 1})_i,
\end{equation}
where the subscript $i=1,2,3$ is the generation index. In other
words, we suppose that the SM quarks and  leptons are in
non-trivial representations of the first factor $SU(5)'$ and
singlets under $SU(5)''$. Notice in Eq.~(\ref{eq:fermions}), that
we have completed  each generation by one ``right-handed'' ({\it
i.e.}, SM singlet) neutrino required to obtain small neutrino
masses via the type-I seesaw mechanism \cite{typeI}. Since the
down quark and charged lepton masses can thus only emerge from
non--renormalizable operators, this may provide a reason why the
down quarks and charged leptons are generally lighter than the up
quarks. Apart from this generic property, however, it would be
desirable to have in this model a more complete understanding of
the observed masses and mixings of quarks and leptons. For this
purpose, we will in the next section attempt to associate the
observed fermion masses and mixing angles with the coupling of the
Higgs and matter fields in Eqs.~(\ref{eq:Higgses}) and
(\ref{eq:fermions}) to the theory space of a deconstructed $U(1)$
symmetry.

\subsection{$U(1)$ theory space}\label{sec:U1theoryspace}
\begin{figure}
\begin{center}
\includegraphics*[bb = 183 521 428 765,height=6.5cm]{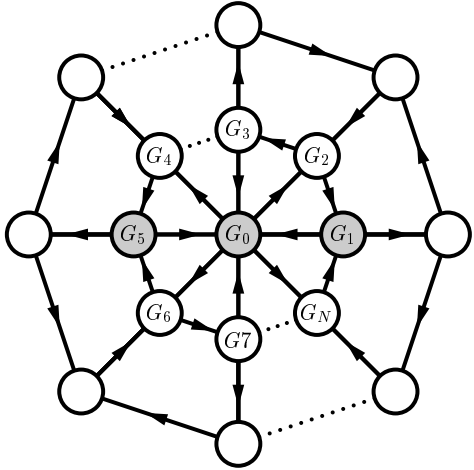}
\end{center}
\vspace*{-6mm} \caption{\small{Spider web theory space for the
deconstructed $U(1)$ gauge theory. Each point or site $i$, where
$i=0,1,\ldots, 2N$, is associated with a gauge group $G_i\equiv
U(1)_i$. An arrow which connects two groups $G_i$ and $G_j$ and
points from $i$ to $j$ denotes a single chiral link superfield
$\phi_{i,j}$ that is charged under $G_i\times G_j$ as $(+1,-1)$
and is a singlet under all the other gauge groups. The first,
second and third generations are placed on the the sites
corresponding to $G_1$, $G_5$, and $G_0$, respectively (gray
circles). For $N$ even, the disk is fitted together by triangular
plaquettes with alternating orientations. The dotted lines
represent possible insertions of extra $U(1)_i$ gauge
groups.}}\label{fig:disk}
\end{figure}

To address the fermion mass hierarchy in the model reviewed in
Sec.~\ref{sec:23splitting}, we will assume that the matter fields
``live'' in a $U(1)$ product group theory space, which describes a
deconstructed manifold. The fermion mass hierarchy arises from
placing the different generations in Eq.~(\ref{eq:fermions}) on
distinct points in the deconstructed manifold. Although there may
be many possibilities, we will first confine ourselves to a theory
space, which is topologically a two--dimensional disk. The reason
for our choice is that a supersymmetry breaking mechanism can be
made readily available in such a theory space
\cite{Arkani-Hamed:2001ed}. We comment on a possible
implementation of this idea in our model at the end of this
section. Our deconstructed manifold is conveniently represented by
the ``moose'' \cite{geor86} or ``quiver'' \cite{douglas:1996xx}
diagram in Fig.~\ref{fig:disk}, which describes a spider web theory space, that is topologically equivalent with a two-dimensional disk.
The center of the spider web theory space is surrounded by $k$ concentric
circles. Each such circle is defined by $N$ sites and  each site $i$, where
$i=0,1,2,\ldots ,kN$, symbolizes one $G_i\equiv U(1)_i$ gauge group.
The total gauge group of our model is therefore $SU(5)'\times
SU(5)''\times U(1)^{kN+1}$ where
$U(1)^{kN+1}\equiv\Pi_{i=0}^{kN}U(1)_i$.
For definiteness, we have in Fig.~\ref{fig:disk} depicted the case $k=2$ and have explicitly labeled only the sites in the inner part of the disk.
When compactifying the
disk later on the real projective plane $RP^2$, we will require
that $N=4m$, where $m$ is some integer. In our spider web theory space, two neighboring sites are connected by a single directed link. The general
organization of the links and their directions is summarized in
Fig.~\ref{fig:disk} for the example of $k=2$. In Fig.~\ref{fig:disk}, an arrow
connecting two sites $i$ and $j$ that
points from $i$ to $j$ denotes a chiral link superfield
$\phi_{i,j}$, which is charged under $G_i\times G_j$ as $(+1,-1)$.
Under all the other $U(1)$ factors and $SU(5)'\times SU(5)''$,
however, the link fields $\phi_{i,j}$ transform only trivially.

It is important to note in our model that for even $N$, the directions of the link fields in the spider web theory space are such that each small triangular or quadratic plaquette has a definite orientation. Any two neighboring
plaquettes have, consequently, opposite orientations. With a single directed link superfield connecting two neighboring sites, only this kind of
configuration allows to have Wilson-loop-type contributions
(in the sense of usual lattice gauge theory)
to the superpotential from {\it every} plaquette in
Fig.~\ref{fig:disk}. As we
will see later, the directions of the link fields are crucial for
generating a realistic hierarchy of Yukawa couplings.
In what
follows, we are interested in the $D$--flat directions
$|\phi_{0,i}|=|\phi_{i,i+1}|\equiv v$ ($i=1,\ldots ,N$) in the
classical moduli space of vacua: All scalar components of the
chiral link superfields have vacuum expectation values (VEV's)
with a universal magnitude $v$. Such a VEV $v$ breaks the $U(1)$
product gauge group spontaneously down to the diagonal subgroup
$U(1)_{\rm diag}$.  Henceforth, we will refer to the field theory defined by
our spider web theory space also as the ``$U(1)$ theory space''
of our model.

Let us now describe how the three generations are incorporated in
our theory space. We suppose that each generation in
Eq.~(\ref{eq:fermions}) is put on a separate site (see
Fig.~\ref{fig:disk}): the first generation ``lives'' on site $1$,
the second on site $5$, and the third on site $0$ in the center of
the disk.\footnote{Instead of putting the second generation on
site 5, we could also choose any site $i$ on the boundary as long
as $i$ is odd for the desired link direction of $\phi_{i,0}$ and
$i,N-i\geq 5$.} This is achieved by giving the first, second and
third generations nonzero $U(1)$ charges exactly under the gauge
groups $U(1)_1$, $U(1)_5$, and $U(1)_0$, respectively, while we
assume that they are singlets under all the other $U(1)$ factors.
Next, we have to specify on the three sites the $U(1)$ charge
assignment to the matter fields within each generation. We choose
the $U(1)$ charges for the fermions in each generation to be
compatible with $SO(10)$ as follows
\begin{subequations}\label{eq:branchings}
\begin{eqnarray}
 SO(10)\supset SU(5)'\times U(1)_1 &:& {\bf 16}_1={\bf 10}(-1)_{1}
+\overline{\bf 5}(3)_1+{\bf 1}(-5)_1,\label{eq:U1}\\
SO(10)\supset SU(5)'\times U(1)_5 &:& {\bf 16}_2={\bf 10}(1)_{2}
+\overline{\bf 5}(-3)_2+{\bf 1}(5)_2,\label{eq:U5}\\
SO(10)\supset SU(5)'\times U(1)_0 &:& {\bf 16}_3={\bf 10}(1)_3
+\overline{\bf 5}(-3)_3+{\bf 1}(5)_3\label{eq:U0},
\end{eqnarray}
\end{subequations}
where the parenthesis contains the the corresponding $U(1)_i$ charge of the multiplets and the subscript denotes the generation index. In Eqs.~(\ref{eq:branchings}), we have, as compared to Eq.~(\ref{eq:fermions}), only kept
the transformation properties of the matter fields under
$SU(5)'$ since they all transform trivially under $SU(5)''$. Note
also in Eq.~(\ref{eq:U1}), that we have made use of an overall sign ambiguity
in the branching rule and assumed that the $U(1)_1$ charges of the first generation are ``flipped'' with respect to the corresponding $U(1)_5$ and $U(1)_0$
quantum numbers in Eqs.~(\ref{eq:U5}) and (\ref{eq:U0}). We emphasize that
the two lighter generations are connected to the third generation by link
fields $\phi_{1,0}$ and $\phi_{5,0}$, which point toward the center. This
orientation is crucial for generating realistic fermion masses and mixings.

At this point, it is important to emphasize that we employ in
Eqs.~(\ref{eq:branchings}) the $SO(10)$ branching rules only as a
mere guideline or organizing principle for the $U(1)$ charge
assignment to the multiplets in Eq.~(\ref{eq:fermions}). Our model does
therefore not possess an actual $SO(10)$ gauge symmetry (for recent
flavor discussions in $SO(10)$ models see, {\it e.g.}, Ref.~\cite{SO10}).
Choosing the $SO(10)$ branching rules as a prescription for the
$U(1)$ charge assignment, however, has several attractive
features. One obvious advantage is, {\it e.g.}, that the quark and
lepton sectors are automatically anomaly-free, such that the
discussion of anomalies is restricted to the Higgs and link fields
only.

One major feature of our model is that the fermion mass hierarchy
is due to the ``location'' of the different generations on
distinct points in theory space (up to the overall sign ambiguity
of the $U(1)$ generators [{\it cf.} Eqs.~(\ref{eq:branchings})]).
This is different, {\it e.g.}, from usual anomalous $U(1)$ models
\cite{Ibanez:1994ig} (for recent related works see, {\it e.g.},
Refs.~\cite{Babu:2003zz,Dreiner:2003yr}), where the
fermion mass hierarchy is understood in terms of flavor--dependent
charges under a single $U(1)$. Notice, that the $U(1)$ charge assignment in
Eqs.~(\ref{eq:branchings}) resembles a gauged $B-L$ symmetry
\cite{Marshak:1979fm}, whose deconstruction has been discussed in
Ref.~\cite{Skiba:2002nx}.

Next, let us consider how the Higgs fields in
Eq.~(\ref{eq:Higgses}) couple to the $U(1)$ theory space. The
$U(1)$ charge assignment to the third generation in
Eq.~(\ref{eq:U0}) already fixes the transformation properties of
$({\bf 5},{\bf 1})^H$. Specifically, to obtain a large top Yukawa
coupling in our model, we suppose that $({\bf 5},{\bf 1})^H$
carries a $U(1)_0$ charge $-2$ and is a singlet under all the
other $U(1)$ gauge groups. The Higgs field $({\bf 5},{\bf 1})^H$
is therefore located as a site variable together with the third
generation on the center of the disk. It is interesting to note,
that the $U(1)_0$ charge assignment to ${\bf 5}^H$ becomes
compatible with $SO(10)$ when considering the ${\bf 5}^H$ as part
of the decomposition ${\bf 10}^H={\bf 5}^H(-2)+\overline{\bf
5}^H(2)$ under $SO(10)\supset SU(5)'\times U(1)_0$, which is also
in agreement with the choice of the $U(1)_0$ generator in
Eq.~(\ref{eq:U0}). All matter and Higgs superfields which are
located on the disk have in common, that they are singlets under
the second factor $SU(5)''$. In contrast to this, we assume that
$({\bf 1},\overline{\bf 5})^H$ in Eq.~(\ref{eq:Higgses}), which is
the only non-trivial $SU(5)''$ representation, carries no $U(1)$
charge at all and is thus not part of the $U(1)$ theory space.

In order to break $U(1)_{\rm diag}$, which is not observed at low energies,
we assume a single vectorlike pair of chiral superfields
$f$ and $\overline{f}$, which resides on the center of the disk. The fields $f$ and $\overline{f}$ are charged under $U(1)_0$ as $+1$ and
 $-1$, respectively, and are singlets under all the other
$U(1)$ and $SU(5)$ gauge groups. In what follows, we will suppose that
the scalar components of the fields $f$ and $\overline{f}$ acquire VEV's
$\langle f\rangle\simeq\langle\overline{f}\rangle\simeq\langle\phi_{i,j}
\rangle\simeq v$, {\it i.e.}, it is assumed that all
$U(1)_i$ symmetries including $U(1)_{\rm diag}$ are broken around
the same scale $v$.

As mentioned earlier, we have an interesting possibility of
supersymmetry breaking in spider web theory space. Supersymmetry
breaking can be implemented in a number of different ways for our
case. Among these we find the mechanism discussed in
Ref.~\cite{Arkani-Hamed:2001ed} to be attractive and unique in
deconstruction. In the remainder of this section, we will briefly
comment on this mechanism.

In Ref.~\cite{Arkani-Hamed:2001ed}, different types of theory
space are shown to preserve supersymmetry only locally, viz., the
interactions on each plaquette are manifestly supersymmetric. If,
however, the topology of theory space has a nontrivial first
homology group, supersymmetry breaking can be seen as a
topological effect. A deconstructed manifold with this property
can, {\it e.g.}, be obtained from the disk in Fig.~\ref{fig:disk},
when we identify diametrically opposite sites and links on the
boundary, which yields a real projective plane $RP^2$ with first
homology group $Z_2$ (this requires in our case $N=4m$, where $m$
is some integer). The phase differences between the gauge
couplings $g_i$ associated with the gauge groups $U(1)_i$ and the
corresponding gauge-Yukawa couplings $h_i=g_ie^{i\theta_i}$ for
the interaction $\sim h_i\psi^\dagger \lambda_i \phi$ (where
$\psi$ and $\phi$ denote the fermionic and scalar components of a
link field connected to the site $i$ with gaugino $\lambda_i$),
can be removed separately in each plaquette by field
redefinitions. In this sense, supersymmetry is preserved
``locally''. Globally, however, there can remain one phase in the
product of all the couplings $h_i$, which cannot be rotated away.
On $RP^2$, this phase is either $+1$, which will lead to exact
global supersymmetry, or $-1$ for maximal supersymmetry breaking
\cite{Arkani-Hamed:2001ed}. The phase being $-1$ rather than
arbitrary, as it would be the case on a circle, can be considered
as an advantage of the spider-web theory space. The supersymmetry
breaking effects are suppressed by a factor $m_{\rm SUSY}^2\sim\Pi_i
g_i^2/(4\pi)^2v^2$ (where $i$ runs over half of the boundary of the
disk in Fig.~\ref{fig:disk})
 due to a number of $N$ loops to account for the nontrivial global twist of
$RP^2$, which can easily produce a TeV scale supersymmetric spectrum.
With this mechanism, one can now address both the fermion mass hierarchy and
supersymmetry breaking in the same theory space.

\section{Quark and lepton masses}\label{sec:quarksandleptons}
\subsection{Up quark sector}\label{sec:upsector}
With the representation content outlined in
Sec.~\ref{sec:U(1)}, we are now in a position to determine
the fermion masses in our model.
Let us first consider the up quark sector. In the notation of Eqs.~(\ref{eq:Higgses}) and (\ref{eq:fermions}), the up quark Yukawa couplings arise from $G$-invariant terms of the type
$\sim({\bf 5},{\bf 1})^H({\bf 10},{\bf 1})_i({\bf 10},{\bf 1})_j$
in the superpotential. Depending on the location of the $({\bf 10},{\bf 1})_i$ matter multiplets on the disk, these terms may be renormalizable or non--renormalizable.
The mass of the top-quark,
{\it e.g.}, emerges from the gauge-invariant renormalizable
operator
$({\bf 5},{\bf 1})^H({\bf 10},{\bf 1})_3({\bf 10},{\bf 1})_3$, with a top Yukawa coupling of order one. This coupling is
renormalizable because the third generation is situated together with
$({\bf 5},{\bf 1})^H$ on the center of the disk carrying $SO(10)$ compatible
$U(1)_0$ charges.

Since the first two generations are located at some distance away
from the center, gauge-invariance under the deconstructed $U(1)$
requires that all other up quark mass terms come from
non--renormalizable operators involving the link fields
$\phi_{i,j}$, which connect the center with the first two
generations. The associated effective Yukawa couplings will thus
be suppressed by inverse powers of the cutoff scale $\Lambda$ of
the effective theory, thereby producing hierarchical mass and
mixing parameters in the fermion sectors. In writing down the
Yukawa couplings, it is of great importance that we work in a
supersymmetric model, where the particular directions of the link
fields as defined in Fig.~(\ref{fig:disk}) constrain the allowed
renormalizable and non--renormalizable terms due to the
holomorphicity of the superpotential. The charm quark mass, {\it
e.g.}, arises dominantly from a non--renormalizable dimension-eight
operator of the type $\Lambda^{-4}\phi_{0,4}^2\phi_{4,5}^2({\bf
5},{\bf 1})^H({\bf 10},{\bf 1})_2 ({\bf 10},{\bf 1})_2$, which
involves two powers of the link fields $\phi_{0,4}$ and
$\phi_{4,5}$. Here, the product of links $\phi_{0,4}\phi_{4,5}$
connects the second generation with the Higgs $({\bf 5},{\bf
1})^H$ in the center along the shortest ``path'' on the disk
consistent with the holomorphicity of the superpotential.
Similarly, the second and third generations mix via the
dimension-six term $\Lambda^{-2}\phi_{0,4}\phi_{4,5}({\bf 5},{\bf
1})^H({\bf 10},{\bf 1})_3({\bf 10},{\bf 1})_2$ associated with the
same path.

Different from the two heavier generations, the mass and mixing terms of the first generation must originate from $U(1)_{\rm diag}$ violating operators, which
involve the $U(1)_{\rm diag}$-breaking fields $f$ or $\overline{f}$ that live
in the center of the disk. This difference arises because the first generation
 carries, with respect to the heavier two generations, opposite charges under
$U(1)_{\rm diag}$. The up quark mass, {\it e.g.}, is generated by the
non--renormalizable term $\Lambda^{-6}f^4\phi_{1,0}^2
({\bf 5},{\bf 1})^H({\bf 10},{\bf 1})_1({\bf 10},{\bf 1})_1$, involving four powers of $f$. This operator contains also two powers of the link field
$\phi_{1,0}$, which connects the first generation with the center. The link field $\phi_{1,0}$ appears therefore also in the operator
$\Lambda^{-3}f^2\phi_{1,0}({\bf 5},{\bf 1})^H({\bf 10},{\bf 1})_3
({\bf 10},{\bf 1})_1$, which mixes the up with the top quark. Correspondingly, the term
$\Lambda^{-5}f^2\phi_{1,0}\phi_{0,4}\phi_{4,5}({\bf 5},{\bf 1})^H
({\bf 10},{\bf 1})_1({\bf 10},{\bf 1})_2$ is responsible for the mixing of the up quark with the charm quark. This operator contains the product of
links $\phi_{1,0}\phi_{0,4}\phi_{4,5}$, which represents on the disk
the shortest connection via holomorphic couplings between the up quark and
the charm quark.

In total, the most general gauge-invariant superpotential containing the
renormalizable and non--renormalizable terms which are relevant for the up
quark masses therefore reads

\begin{eqnarray}
 \mathcal{W}&\supset&({\bf 5},{\bf 1})^H({\bf 10},{\bf 1})_3
({\bf 10},{\bf 1})_3+\frac{\phi_{0,4}^2\phi_{4,5}^2}{\Lambda^4}
({\bf 5},{\bf 1})^H({\bf 10},{\bf 1})_2({\bf 10},{\bf 1})_2\nonumber\\
&+&\frac{f^4\phi_{1,0}^2}{\Lambda^6}
({\bf 5},{\bf 1})^H({\bf 10},{\bf 1})_1({\bf 10},{\bf 1})_1
+\frac{\phi_{0,4}\phi_{4,5}}{\Lambda^2}({\bf 5},{\bf 1})^H
({\bf 10},{\bf 1})_3({\bf 10},{\bf 1})_2
\nonumber\\
&+&\frac{f^2\phi_{1,0}}{\Lambda^3}({\bf 5},{\bf 1})^H({\bf 10},{\bf 1})_3
({\bf 10},{\bf 1})_1+\frac{f^2\phi_{1,0}\phi_{0,4}\phi_{4,5}}{\Lambda^5}
({\bf 5},{\bf 1})^H({\bf 10},{\bf 1})_1({\bf 10},{\bf 1})_2+\dots,
\end{eqnarray}
where the dots denote negligible higher-order terms and where we have
not explicitly written the different Yukawa couplings of order one.
When all the link and site fields $\phi_{i,j}$ and $f$
acquire their VEV's around the deconstruction scale $v$, the up
quark mass matrix is then given by the well--known texture
\begin{equation}\label{eq:uptexture}
 M_u=\langle H\rangle
\left(
\begin{matrix}
 \epsilon^6 & \epsilon^5 & \epsilon^3\\
 \epsilon^5 & \epsilon^4 & \epsilon^2\\
 \epsilon^3 & \epsilon^2 & 1
\end{matrix}\right),
\end{equation}
where we have introduced the small symmetry-breaking parameter
$\epsilon\equiv v/\Lambda\simeq 0.2$. Since the texture in
Eq.~(\ref{eq:uptexture}) can already fully account for the observed CKM angles, the down quark mixing should not become too large in order to avoid
conflict with experiment. As we will see in Sec.~\ref{sec:downsector}, the mixing in the down sector does indeed not exceed the up quark mixing.

\subsection{Down quark and charged lepton sector}\label{sec:downsector}
The construction of the down quark and charged lepton Yukawa coupling
terms goes along the same lines as for the up quarks in
Sec.~\ref{sec:upsector}, except for the
difference that the matter fields and $({\bf 1},\overline{\bf 5})^H$
transform in $G$ under different $SU(5)$ factors.
The down quark and charged lepton Yukawa couplings must therefore emerge from
$G$-invariant terms $\sim (\overline{\bf 5},{\bf 5})^H({\bf 1},\overline{\bf 5})^H({\bf 10},{\bf 1})_i(\overline{\bf 5},{\bf 1})_j$, where
$(\overline{\bf 5},{\bf 5})^H$ denotes in the doublet--triplet splitting
mechanism reviewed in Sec.~\ref{sec:23splitting} a linear combination of Higgs
superfields which transforms under $SU(5)'\times SU(5)''$ as a bifundamental
representation. When $(\overline{\bf 5},{\bf 5})^H$ acquires
its VEV at the GUT scale, $G\times F$ is broken down to the low-energy subgroup
$G_{SM}\times F'$ (see Refs.~\cite{Witten:2001bf, Barbieri:1994jq}).

To generate the down quark and charged lepton masses, we assume two such
bifundamental Higgs superfields
$\Phi_+$ and $\Phi_-$, which are put as site variables
on the center of the disk and transform under
$SU(5)'\times SU(5)''\times U(1)_0$ as
$\Phi_+\sim (\overline{\bf 5},{\bf 5})^H(+2)$ and $\Phi_-\sim (\overline{\bf 5},{\bf 5})^H(-2)$, where in each of the last two expression the second
parenthesis contains the value of the $U(1)_0$ charge. Under all the other $U(1)_i$ groups,
$\Phi_+$ and $\Phi_-$ transform trivially. By the same arguments as in Sec.~\ref{sec:upsector},
we then find for the relevant superpotential terms responsible for
the down quark charged lepton masses
\begin{eqnarray}\label{eq:downYukawas}
\mathcal{W}&\supset&
\frac{\Phi_+}{\Lambda'}
({\bf 1},\overline{{\bf 5}})^H({\bf 10},{\bf 1})_3
(\overline{{\bf 5}},{\bf 1})_3
+\frac{\phi_{5,0}^2\Phi_+}{\Lambda^2\Lambda'}({\bf 1},\overline{\bf 5})^H
({\bf 10},{\bf 1})_2(\overline{\bf 5},{\bf 1})_2\nonumber\\
&+&\frac{\phi_{0,2}^2\phi_{2,1}^2\Phi_-}{\Lambda^4\Lambda'}({\bf 1},\overline{{\bf 5}})^H({\bf 10},{\bf 1})_1
(\overline{\bf 5},{\bf 1})_1+\dots,
\end{eqnarray}
where the dots denote irrelevant higher-order operators and where we have not
explicitly written the Yukawa couplings of order one. In Eq.~(\ref{eq:downYukawas}), the scale $\Lambda'$ is related to the GUT--scale $M_{\rm GUT}\simeq 10^{16}\:{\rm GeV}$ by $\Lambda'\simeq M_{\rm GUT}/\epsilon'$, where $\epsilon'\sim 0.1$. Observe that $\Lambda'$ is a common factor to the down sector parameterizing $\textrm{tan}\:\beta$ and thus plays no role for the flavor structure.
When $\Phi_+$ and $\Phi_-$ acquire similar VEV's
$\langle \Phi_+\rangle\simeq\langle\Phi_-\rangle\sim M_{\rm GUT}$, the mass
matrix of the down quarks and charged leptons takes the form
\begin{equation}\label{eq:downquarkmatrix}
 M_d=\epsilon'\langle \tilde{H}\rangle
\left(
\begin{matrix}
 \epsilon^4 & \epsilon^6 & \epsilon^3\\
 \epsilon^{10} & \epsilon^2 & \epsilon^2\\
 \epsilon^8 & \epsilon^3 & 1
\end{matrix}\right),
\end{equation}
where the rows and columns are spanned by the $({\bf 10},{\bf 1})_i$ and
$(\overline{\bf 5},{\bf 1})_j$, respectively, and where we have a moderate
${\rm tan}\beta\equiv\langle H\rangle/\langle\tilde{H}\rangle\sim
 10$. In total, one therefore obtains for the quark and
charged lepton mass ratios
\begin{subequations}\label{eq:massratiosandCKM}
\begin{eqnarray}
  m_u:m_c:m_t&=&\epsilon^6:\epsilon^4:1,\\
 m_d:m_s:m_b&=&\epsilon^4:\epsilon^2:1,\\
 m_e:m_\mu:m_\tau&=&\epsilon^4:\epsilon^2:1.
\end{eqnarray}
The CKM angles are of the orders
\begin{equation}
 V_{us}\sim\epsilon,\quad V_{cb}\sim\epsilon^2,\quad
V_{ub}\sim\epsilon^3.
\end{equation}
\end{subequations}
In Eq.~(\ref{eq:downquarkmatrix}), we observe that the charged
lepton mixing angles are $\lesssim \epsilon^3$. The large leptonic
mixing angles must therefore be almost entirely generated in the
neutrino sector. The neutrino masses and mixing angles will be
discussed now.

\subsection{Neutrino masses}\label{sec:neutrinos}

Following the generic approach put forward in
Ref.~\cite{Balaji:2003st}, we shall relate the absolute neutrino
mass scale to the deconstruction scale via a dynamical realization
of the type-I seesaw mechanism, where the inverse lattice spacing
$\sim v$ is identified with the usual $B-L$ breaking scale
$v\simeq 10^{14}\:{\rm GeV}$. To leading order, the total
effective $3\times 3$ neutrino mass matrix $M_\nu$ can thus be
written as $M_\nu=-M_DM_R^{-1}M_D^T$, where, as usually, $M_D$
denotes the Dirac neutrino mass matrix and $M_R$ is the
right-handed Majorana mass matrix. The qualitative difference
between $M_D$ and $M_R$ is, of course, that $M_D$ is protected by
$G_{SM}$ down to the electroweak scale, while $M_R$ can already
emerge at the deconstruction scale $v$ through the Yukawa
interactions between the right-handed neutrinos (which are
vectorial with respect to $G$) and the link fields.

When determining $M_D$ and $M_R$ in the same way like $M_u$ and
$M_d$ in Secs.~\ref{sec:upsector} and \ref{sec:downsector},
however, we find that the minimal theory space introduced in
Sec.~\ref{sec:U1theoryspace} would only give small neutrino mixing
angles. To arrive at a large neutrino mixing, one may deviate from
minimality and add extra link fields to our $U(1)$ theory
space. Specifically, we assume that each of the directed link
superfields $\phi_{i,j}$ defined in Sec.~\ref{sec:U1theoryspace}
is accompanied by a pair of vectorlike chiral link superfields
$\chi_{i,j}$ and $\overline{\chi}_{j,i}$ which point into opposite
directions and acquire universal VEV's of the order
$\langle\chi_{i,j}\rangle\simeq\langle\overline{\chi}_{j,i}\rangle\simeq
v$. While $\phi_{i,j}$ carries the $G_i\times G_j$ charges
$(+1,-1)$, the fields $\chi_{i,j}$ and $\overline{\chi}_{j,i}$ are
charged under $G_i\times G_j$ as $(+8,-8)$ and $(-8,+8)$,
respectively. One can check that the incorporation of the link
fields $\chi_{i,j}$ and $\overline{\chi}_{j,i}$ has no effect on
our results in Sec.~\ref{sec:upsector} and \ref{sec:downsector}
for the charged fermion mass ratios and CKM angles summarized in
Eqs.~(\ref{eq:massratiosandCKM}). In contrast to this, the extra
Yukawa interactions between the $\overline{\chi}_{j,i}$ and the
right-handed neutrinos introduce a large off-diagonal term in $M_R$,
which results in a large
atmospheric neutrino mixing angle $\theta_{23}\sim 1$. This is a
generalization of the scenario for soft breaking of the
$L_e-L_\mu-L_\tau$ lepton number in the right-handed sector
\cite{Grimus:2001ex}.

A fully realistic description of bilarge neutrino mixing with normal neutrino mass hierarchy can then be
obtained by adding on the sites
extra Higgs superfields known from standard realizations
of the seesaw mechanism. For example, we can assume an $SU(5)'\times SU(5)''$
singlet Higgs superfield  $({\bf 1},{\bf 1})^H$, which is placed together with the second generation on the site 5. The $({\bf 1},{\bf 1})^H$ carries a charge $-10$ under $U(1)_5$ and is a singlet under all the other
$U(1)_i$ groups. This $U(1)_5$ charge
assignment becomes compatible with $SO(10)$ on the site 5, when
we identify $({\bf 1},{\bf 1})^H$ with the $SU(5)'$ singlet in the decomposition
\begin{eqnarray}\label{eq:126}
 SO(10)\supset SU(5)'\times U(1)_5 &:&
\overline{\bf 126}^H={\bf 1}^H(-10)+{\bf 15}^H(6)+\dots,
\end{eqnarray}
where we have only written the subrepresentations relevant for $M_\nu$. The
$({\bf 1},{\bf 1})^H$ couples to the right-handed
neutrinos via a renormalizable term
$({\bf 1},{\bf 1})^H({\bf 1},{\bf 1})_2({\bf 1},{\bf 1})_2$, thereby
supplementing $M_R$ with an additional parameter.
Choosing $\langle({\bf 1},{\bf 1})^H\rangle\simeq \epsilon^7 v$, the effective
neutrino mass matrix comes to the familiar form
\begin{equation}\label{eq:Meff}
 M_\nu=\epsilon\frac{\langle H\rangle^2}{v}
\left(
\begin{matrix}
 \epsilon^4 & \epsilon^2 & \epsilon\\
\epsilon ^2 & 1 & 1\\
\epsilon & 1 & 1
\end{matrix}
\right).
\end{equation}
Taking $\sim 4\times 10^{-2}\:{\rm eV}$ as the heaviest active neutrino mass,
we find from Eq.~(\ref{eq:Meff}) for the deconstruction scale
a value $v\simeq 10^{14}\:{\rm GeV}$, which is of the order the usual
$B-L$ breaking scale. In its present form, however, the 2-3 subblock of
$M_\nu$ in Eq.~(\ref{eq:Meff}) has a determinant which is much smaller than
$\epsilon$, so that the solar mixing angle would be close to maximal.

In order to obtain a large but not maximal solar mixing angle, we can invoke
the type-II seesaw mechanism \cite{typeII}, which provides a contribution of order $\sim\epsilon$ to the 2-3 subblock of $M_\nu$
in Eq.~(\ref{eq:Meff}), thus suppressing $\theta_{23}$ down to values
$\sim \pi/6$. The type-II seesaw mechanism
may be implemented in our model by adding on the
center of the disk a pair of conjugate Higgs superfields as site variables,
that transform under $SU(5)'\times SU(5)''\times U(1)_0$ as
$({\bf 15},{\bf 1})^H(6)$ and $(\overline{\bf 15},{\bf 1})^H(-6)$,
respectively, but which are singlets under all the other $U(1)_i$ gauge
groups (for a discussion of phenomenological implications see, {\it e.g.}, Ref.~\cite{Rossi:2002zb}). The $U(1)_0$ charges of these Higgs fields are $SO(10)$ compatible,
as can be seen from the branching rule in
Eq.~(\ref{eq:126}), by replacing the gauge group $U(1)_5$ by $U(1)_0$.
The superpotential couplings for the type-II
seesaw mechanism involve a renormalizable term
$M_{15}({\bf 15},{\bf 1})^H(\overline{\bf 15},{\bf 1})^H$,
where $M_{15}$ is some high mass scale. Now, after integrating out the
heavy Higgs fields, the contribution to the 3-3 element of
$M_\nu$ in Eq.~(\ref{eq:Meff}) is of the order $\sim\epsilon^9
\langle H\rangle^2/M_{15}$. If $M_{15}\simeq 10^{9}\:{\rm GeV}$, then
the total effective neutrino mass matrix $M_\nu$ can assume a similar form
like in Eq.~(\ref{eq:Meff}), with the difference that the determinant of the
2-3 subblock is now of the order $\sim \epsilon$.  For our choice of parameters, the model can thus lead to a normal active neutrino mass hierarchy
\begin{eqnarray}
 m_1:m_2:m_3 &=& \epsilon : \epsilon : 1,
\end{eqnarray}
where $m_1$, $m_2$, and $m_3$ are the active neutrino masses with
solar and atmospheric mass squared differences of the orders
$\Delta m_{\odot}^2\simeq 10^{-4}\:{\rm eV}^2$ and $\Delta m_{\rm
atm}^2\simeq 10^{-3}\:{\rm eV}^2$, respectively. In this case, we
then have a small reactor angle $\theta_{13}\sim \epsilon$, a
large but not maximal solar angle $\theta_{12}\sim 1$ and a large
atmospheric angle $\theta_{23}\sim 1$, which can be maximal. Our
model can therefore accommodate current global fits to neutrino
oscillation data \cite{Maltoni:2004ei}.

\section{Anomaly Cancelation}\label{sec:anomalies}
Although the $SO(10)$ compatible $U(1)$ charge assignment to the
fermions in Eqs.~(\ref{eq:branchings}) is anomaly-free, the Higgs
field sector in its present form would contain anomalies. Note
that, in our spider web theory space, any Higgs superfield with
anomalous coupling is either a link field $\phi_{i,j}$ or must be
situated as a site variable on a single site. The anomalies coming
from these site variables may be directly canceled by simply
adding in a standard fashion extra fields on the sites where they
reside. In contrast to this, we shall now consider the possibility
to cancel the pure and mixed anomalies associated with the link
fields $\phi_{i,j}$ without introducing any new fields in the
low-energy effective theory.

First, we will discuss the pure ({\it i.e.}, non-mixed) and gauge-gravitational
anomalies. When the topology of the spider web theory space in Fig.~\ref{fig:disk} is that
of a disk, the link fields $\phi_{i,j}$ would give rise to pure and gauge-gravitational anomalies on each site of the boundary. Interestingly, these anomalies are completely eliminated, when the spider web theory space is, instead, compactified on the real projective plane $RP^2$.
Observe that the removal of the pure and gauge-gravitational anomalies
by compactifying on $RP^2$ relies in an essential way on our requirement to
have a definite orientation for each small plaquette in Fig.~\ref{fig:disk}.
The compactification on $RP^2$ alone, however, does not remove the mixed anomalies induced by the link fields.

\begin{figure}
\begin{center}
\includegraphics*[bb = 235 575 375 716,height=4.0cm]{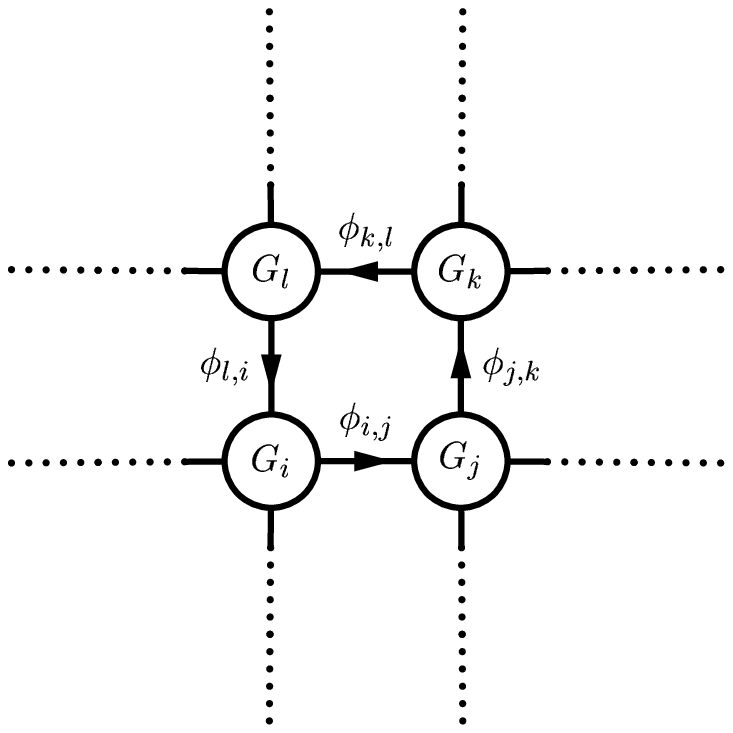}
\end{center}
\vspace*{-7mm} \caption{\small{Plaquette in the spider web theory
space, considered for the Chern--Simons terms.}}
\label{fig:plaquette}
\end{figure}

Now, let us discuss the cancelation of the mixed gauge anomalies. To this end,
we add Wess--Zumino (WZ) terms \cite{Wess:1971yu}, which can be viewed as
emerging from integrating out heavy fermions with large Yukawa couplings.
 The mass scale of these extra fermions is
one or two orders of magnitude above the inverse lattice spacing
$v\sim 10^{14}\:{\rm GeV}$. In doing so, we follow
Refs.~\cite{Dudas:2003iq,Falkowski:2002vc}, wherein the case of a deconstructed fifth
dimension has been analyzed. Let us consider in Fig.~\ref{fig:disk} a site $i\neq 0$ which is not in the center (a similar argumentation holds for $i=0$). The site $i$ is connected to its
four neighboring sites $j_1$, $j_2$, $j_3$, and $j_4$ by the link fields
$\phi_{i,j_1}$, $\phi_{i,j_2}$, $\phi_{j_3,i}$, and $\phi_{j_4,i}$. Note that
$\phi_{i,j_1}$ and $\phi_{i,j_2}$ point from $i$ to $j_1$ and $j_2$, while $\phi_{j_3,i}$ and $\phi_{j_4,i}$ point from $j_3$ and $j_4$
toward the site $i$. The directions of the link fields are a
result of the property of our theory space, that two neighboring small plaquettes have alternating orientations. Under an infinitesimal chiral
gauge transformation on the site $i$, the vector multiplet $V_i$ belonging to the gauge group $U(1)_i$ transforms as $V_i\rightarrow V_i+{\rm i}(\overline{\Lambda}_i-\Lambda_i)$, where $\Lambda_i$ is the gauge parameter.
Denoting by $j^\mu_i$ the chiral current associated with the gauge
transformation at the site $i$, we can arrange in the one-loop 3-point function
$\langle 0|Tj^\mu_i j^\nu_k j^\rho_l|0\rangle$ the anomalies symmetrically
among the three involved currents.
In a superfield language, the anomalous variation of the link field Lagrangian $\mathcal{L}_{\rm link}$ corresponding to the gauge transformation $\Lambda_i$ can
then be written as
\begin{eqnarray}\label{eq:anomalies}
 \delta_{\Lambda_i}\mathcal{L}_{\rm link}&=&
-\frac{{\rm i}}{12\pi^2}
\int d^2\theta\:\Lambda_i
\left[W^\alpha_{j_1}W_{\alpha,j_1}-2W^\alpha_iW_{\alpha,j_1}+(j_1\leftrightarrow j_2)\right.\nonumber\\
&&\left.-W^\alpha_{j_3}W_{\alpha,j_3}+
2W^\alpha_iW_{\alpha,j_3}+(j_3\leftrightarrow j_4)\right]+{\rm h.c.},
\end{eqnarray}
where $W_{\alpha,i}$ denotes the supersymmetric field strength of the gauge group $U(1)_i$. An analogous expression to
Eq.~(\ref{eq:anomalies}) holds for the mixed anomalies
$\delta_{\Lambda_0}\mathcal{L}_{\rm link}$,
associated with a chiral gauge transformation $\Lambda_0$ on the site 0 in the center of the theory space. The mixed anomalies $\delta_{\Lambda_i}\mathcal{L}_{\rm link}$ can be
canceled in the low-energy effective theory by appropriate WZ terms,
which are constructed from local polynomials in the link
fields $\phi_{i,j}$ and gauge multiplets $V_i$. To remove the mixed
anomalies $\delta_{\Lambda_i}\mathcal{L}_{\rm link}$ in Eq.~(\ref{eq:anomalies}), we add to our model
the WZ terms
\begin{eqnarray}\label{eq:WZ}
\mathcal{L}_{\rm WZ}^i &=&-\frac{1}{24\pi^2}\int
 d^2\theta\left\{
{\rm log}(\phi_{i,j_1}/v)\left[(C_1-1)W^\alpha_iW_{\alpha,i}+
(C_1-1)W^\alpha_{j_1}W_{\alpha,j_1}\right.\right.\nonumber\\
&&\left.+(C_1+2)W^\alpha_iW_{\alpha,j_1}\right]+
{\rm log}(\phi_{j_3,i}/v)\left[(C_1-1)W^\alpha_iW_{\alpha,i}
+(C_1-1)W^\alpha_{j_3}W_{\alpha,j_3}\right.\nonumber\\
&&\left.\left.+(C_1+2)W^\alpha_iW_{\alpha,j_3}\right]\right\}
-\frac{C_2}{24\pi^2}
\int d^4\theta\left[(V_iD_\alpha V_{j_1}-V_{j_1}D_\alpha V_i)
(W^\alpha_i+W^\alpha_{j_1})\right.\nonumber\\
&&\left.(V_iD_\alpha V_{j_1}-V_{j_1}D_\alpha V_i)
(W^\alpha_i+W^\alpha_{j_1})\right]
+(j_1\leftrightarrow j_2)+(j_3\leftrightarrow j_4)+{\rm h.c.},
\end{eqnarray}
where $C_1$ and $C_2$ are some suitable parameters. In Eq.~(\ref{eq:WZ}), the
terms with factors $C_1$ and $C_2$ match in the continuum limit onto six-dimensional (6D)
Chern--Simons couplings, when taking the sum of these operators around a plaquette. To see this, let us consider the quadratic
plaquette shown in Fig.~\ref{fig:plaquette} as a part of the spider web theory space, which is spanned by the sites
$i$, $j$, $k$, and $l$.
From Eq.~(\ref{eq:WZ}), we find that the sum of all terms with
factors $C_1$ and $C_2$, which correspond to the plaquette, is
given by
\begin{eqnarray}\label{eq:plaquetteterm}
 \mathcal{L}_{\rm CS}^{ijkl}&=&
-\frac{C_1}{24\pi^2}
\int d^2\theta\:{\rm log}(\phi_{i,j})\left[
W_{\alpha,i}W_i^\alpha+W_{\alpha,j}W^\alpha_j+W_{\alpha,i}W^\alpha_j
\right]\nonumber\\
&&-\frac{C_2}{24\pi^2}\int d^4\theta
\left[(V_iD_\alpha V_{j}-V_jD_\alpha V_i)(W^\alpha_i+W^\alpha_j)\right]\nonumber\\
&&+((i,j)\leftrightarrow (j,k))+((j,k)\leftrightarrow (k,l))
+((k,l)\leftrightarrow (l,i))+{\rm h.c.},
\end{eqnarray}
where we have indicated in the last line a cyclic permutation of the four
sides of the plaquette. We parameterize the link fields attached to the site $i$ as $\phi_{i,j}=\frac{v}{\sqrt{2}}e^{(\Sigma_{i,j}+{\rm i}G_{i,j})/v}$ and
$\phi_{l,i}=\frac{v}{\sqrt{2}}e^{(\Sigma_{l,i}+{\rm i}G_{l,i})/v}$. In the continuum limit, $G_{i,j}$ and $G_{l,j}$ become $G_{i,j}\rightarrow A_5$ and
$G_{l,i}\rightarrow -A_6$, where $A_5$ and $A_6$ are the 5th and 6th
components of the $U(1)$ gauge field of the 6D theory.
Expanding around the site $i$, the term
$\mathcal{L}^{ijkl}_{\rm CS}$ in Eq.~(\ref{eq:plaquetteterm}) matches in the
continuum limit onto
\begin{equation}
\mathcal{L}^{ijkl}_{\rm CS}\rightarrow-\frac{1}{24\pi^2}\epsilon^{\mu\nu\rho\sigma}\left[3C_1(\partial_4A_5-\partial_5A_4)\partial_\mu A_\nu\partial_\rho
A_\sigma-4C_2\partial_4A_\mu\partial_5 A_\nu\partial_\rho A_\sigma\right],
\end{equation}
which reproduces the 6D Chern--Simons term
$\mathcal{L}_{\rm CS}=-(C_1/8\pi^2)\epsilon^{\alpha\beta\mu\nu\rho\sigma}
\left[\partial_\alpha A_\beta\partial_\mu A_\nu\partial_\rho A_\sigma\right]$
for the choice $C_2=(3/2)C_1$. To determine the constant $C_1$, note in
Eq.~(\ref{eq:WZ}) that the effective moduli fields
 ${\rm log}(\phi_{i,j})$ transform under gauge transformations on the neighboring sites as ${\rm log}(\phi_{i,j})\rightarrow{\rm log}(\phi_{i,j})+
2{\rm i}(\Lambda_i-\Lambda_j)$. As a consequence, the anomalous variation
$\delta_{\Lambda_i}\mathcal{L}^i_{\rm WZ}$ of the WZ term in Eq.~(\ref{eq:WZ})
obeys
$\delta_{\Lambda_i}\mathcal{L}_{\rm WZ}=-\delta_{\Lambda_i}\mathcal{L}_{\rm link}$ and thus cancels the mixed anomalies in Eq.~(\ref{eq:anomalies}) when
$C_1=0$, {\it i.e.}, the Chern--Simons term has to vanish.

The correspondence to the 6D model can be established in the
infrared limit via the tower of gauge bosons generated by the
kinetic term $\sum_{(i,j)}(D_\mu \phi_{i,j})^\dagger D^\mu
\phi_{i,j}$, in which the covariant derivative $D_\mu$ acts on the
scalar components of all the links $\phi_{i,j}$ connecting
neighboring sites $i$ and $j$. At low energies, the associated
gauge boson mass matrix is then schematically given by the
contribution $\sim g^2v^2\sum_{(i,j)}(A_i^\mu-A_j^\mu)^2$ to the
total Lagrangian, where $A_i^\mu$ are the $U(1)_i$ gauge bosons
and where we have, for simplicity, assumed universal gauge
couplings for all the $U(1)$ factors and universal VEV's $v$ of
the link fields. Although our spider--web theory space is
topologically equivalent to the real projective plane $RP^2$, a
variation of the individual gauge couplings and link field VEV's
would offer many possibilities to realize distinct geometries in
the continuum limit. At this level, the identification of the
exact geometry of the corresponding deconstructed manifold,
however, would require further study of the detailed gauge boson
spectrum, which is beyond the scope of this paper. Since in our
particular fermion mass model the universal inverse lattice
spacing is fixed by the neutrino sector to be of the order $v\sim
10^{14}\:\textrm{GeV}$, a straightforward way to arrive at a large
$N$ limit would be to extend the exterior of the disk in
Fig.~\ref{fig:disk} by simply adding extra quadratic plaquettes to
the boundary.

Generally, we wish to point out that the cancelation of the pure
and gauge--gravitational anomalies allows also different ways to
organize the orientation of the link fields. Consider, for
instance, a compactification of the the spider--web theory space
on $RP^n$ with $n>2$ and even, by partitioning the boundary into
$n$ equivalent line segments that are identified under a rotation
of the disk by an angle $2\pi/n$. We can arrive in this case at a
similar phenomenology of fermion masses, when in
Fig.~\ref{fig:disk} all links on the concentric circles point,
{\it e.g.}, in clockwise direction while all the radial links
within each of the $n$ segments of the disk point either inward or
outward but alternate between two adjacent segments.

Finally, one can ask whether our model is embedable into a string
construction as suggested by quiver gauge theories
\cite{douglas:1996xx, Cachazo:2001gh}. Recent advances in the
AdS/CFT correspondence \cite{Gauntlett:2004yd} show that infinite
classes of 4D $\mathcal{N}=1$ SUSY quiver gauge theories can be
viewed as a stack of D3-branes probing a singular Calabi--Yau
manifold. For our specific model, however, such embeddings remain
to be seen.

\section{Conclusions}\label{sec:conclusions}

In this paper, we have presented a model, wherein the observed fermion masses
and mixing angles emerge from a deconstructed $U(1)$ theory space. We
have extended a supersymmetric $SU(5)'\times SU(5)''$ product GUT,
which has been previously suggested for solving the
doublet--triplet splitting problem
\cite{Witten:2001bf,Barbieri:1994jq}, by a deconstructed $U(1)$
theory space with disk structure. The different generations of the
SM fermions live at different sites of the disk. Upon breaking the
$U(1)$ product group by the link fields around the $B-L$ breaking scale
$v\simeq10^{14}$ GeV, the effective Yukawa couplings and mixing matrices of
the fermions are correctly reproduced through non--renormalizable operators.
The $U(1)$ charge assignment to the fermions is compatible with $SO(10)$
and, thus, free from anomalies. This is a major
difference compared to usual, {\it e.g.}, anomalous $U(1)$ models,
where the SM generations differ by flavor--dependent charges, which
appears to be somewhat adhoc from a bottom--up point of view. The
neutrino mass matrix receives contributions from both type--I and
type--II seesaw mechanisms. Among many possibilities, we have
advocated the supersymmetry breaking scenario of
Ref.~\cite{Arkani-Hamed:2001ed}, which is unique to deconstructed
models. To do so, the original disk theory space is thought to be part
of a larger structure, viz.,  a spider web theory space. When
diametrically opposite sites and links on the boundary of this
space are identified, we arrive at an $RP^2$ manifold with nontrivial
first homology group $Z_2$. The interactions on each plaquette are here
required to be manifestly supersymmetric. The nontrivial global
twist of $RP^2$ can be viewed as the source of supersymmetry breaking. Thus,
both the fermion mass matrix structures and supersymmetry breaking
can now be addressed in the same theory space, which we find
interesting and economic. The choice of the charges for the link
fields, which defines a direction for the links connecting two
sites, is such that neighboring plaquettes have alternating
orientations. As a consequence, all the sites have the same number of
``ingoing" and ``outgoing" link fields. This arrangement insures
that the contributions to the pure and gravitational anomalies on
each site vanish automatically. We cancel the mixed anomalies,
along the line of Refs.~\cite{Dudas:2003iq,Falkowski:2002vc}, by
Wess--Zumino terms, which can be considered as a
result of integrating out heavy fermions with masses one or two
orders of magnitude above the $B-L$ breaking scale. We have
examined possible Chern--Simons terms on a rectangular plaquette,
which nevertheless do not play a role in the anomaly cancelations,
and have shown that they have a correct 6D continuum limit.

It would be clearly interesting to develop descriptions of our model based on
gauge groups like $SO(10)$ or $E_6$ with a universal GUT/deconstruction scale. In possible variations of our model one could (as proposed,
{\it e.g.}, in Ref.~\cite{Skiba:2002nx}) also think of shifting anomalies
between the gauge groups using a deconstructed version of the anomaly inflow
mechanism known from string theory \cite{Callan:1984sa}.

\section*{Acknowledgments}
We would like to thank K.S. Babu, S. Pokorski, and S. Vempati for useful comments and discussions.
One of us (G.S.) would like to thank the Fermilab Theory Group for their warm
hospitality and support during the stay as a summer visitor, where part of
this work was developed. This work was supported by the U.S. Department
of Energy under Grant Numbers DE-FG02-04ER46140 and DE-FG02-04ER41306
(T.E. and G.S.).

\end{document}